\documentstyle[11pt]{article}

\newcommand{\half}{{\scriptstyle{\frac{1}{2}}}}
\newcommand{\LP}{\lambda \Phi^4}
\newcommand{\dl}{\delta^{(3)}({\bf r})}
\newcommand{\del}{{\mbox{\boldmath $\nabla$}}}
\newcommand{\BE}{\begin{equation}}
\newcommand{\EE}{\end{equation}}
\newcommand{\BA}{\begin{eqnarray}}
\newcommand{\EA}{\end{eqnarray}}
\newcommand{\vol}{{\sf V}}
\newcommand{\num}{{\sf N}}
\begin{document}
\begin{titlepage}
\begin{flushright}
{ \small DE-FG05-92ER41031-50 }
\end{flushright}

\vspace*{1mm}
\begin{center}

            {\LARGE{\bf Physical mechanisms generating \\ 
                 spontaneous symmetry breaking  \\
\vspace*{4mm}
                  and a hierarchy of scales}}

\vspace*{11mm}
{\Large  M. Consoli}
\vspace*{2mm}\\
{\large
Istituto Nazionale di Fisica Nucleare, Sezione di Catania \\
Corso Italia 57, 95129 Catania, Italy}
\vspace*{2mm}\\
and \\
\vspace*{2mm}
{\Large P. M. Stevenson}
\vspace*{2mm}\\
{\large T. W. Bonner Laboratory, Physics Department \\
Rice University, P.O. Box 1892, Houston, TX 77251-1892, USA}
\vspace{7mm}\\
\end{center}
\begin{center}
{\bf Abstract}
\end{center}

  We discuss the phase transition in $3+1$~dimensional $\LP$ theory 
from a very physical perspective.  The particles of the symmetric phase 
(`phions') interact via a hard-core repulsion and an induced, long-range 
$-1/r^3$ attraction.  If the phion mass is sufficiently small, the 
lowest-energy state is not the `empty' state with no phions, but is a 
state with a non-zero density of phions Bose-Einstein condensed in the 
zero-momentum mode.  The condensate corresponds to the 
spontaneous-symmetry-breaking vacuum with $\langle \Phi \rangle \neq 0$ 
and its excitations (``phonons'' in atomic-physics language) correspond 
to Higgs particles.  The phase transition happens when the phion's physical 
mass $m$ is still positive; it does not wait until $m^2$ passes through 
zero and becomes negative.  However, at and near the phase transition, 
$m$ is much, much less than the Higgs mass $M_h$.  This interesting physics 
coexists with ``triviality;'' all scattering amplitudes vanish in the 
continuum limit, but the vacuum condensate becomes infinitely dense.  
The ratio ${{m}\over{M_h}}$, which goes to zero in the continuum limit, 
can be viewed as a measure of non-locality in the regularized theory.  
An intricate hierarchy of length scales naturally arises.  We speculate 
about the possible implications of these ideas for gravity and inflation.  


\end{titlepage}
 
\setcounter{page}{1}

\vfill
\eject
\setcounter{equation}{0}
\section{Introduction} 

    Spontaneous symmetry breaking is an essential component of current 
theories of particle physics.   All particles in the Standard Model 
acquire their masses from a non-vanishing expectation value 
$\langle \Phi \rangle\neq 0 $ of a self-interacting scalar field.  
The idea is simple and has a long history, so one might think that 
little remains to be understood.  However, a basic question remains 
to be settled: the nature of the phase transition in the $\LP$ scalar 
field theory.  

    At the classical level one need only look at the potential 
\BE
\label{clpot}
    V_{\rm cl}(\phi)= {{1}\over{2}} m^2 \phi^2 + 
{{\lambda}\over{4!}}\phi^4  
\EE
to see that the phase transition, as one varies the $m^2$ parameter, is 
second order and occurs at $m^2=0$.  In the quantum theory, however, the 
question is more subtle.  Clearly, the symmetric vacuum is locally 
stable if its excitations have a physical mass $m^2 >0$ and locally unstable 
if $m^2 <0$.  However, there remains the question of whether an $m^2>0$ 
symmetric vacuum is necessarily {\it globally} stable.  Could the phase 
transition actually be first order, occurring at some small but positive 
$m^2$?  

    The standard approximation methods for the quantum effective potential 
$V_{\rm eff}(\phi)$ give contradictory results on this crucial issue 
\cite{cw}.  The straightforward one-loop approximation predicts a first-order 
transition occurring at a small but positive value of the physical 
(renormalized) mass squared, $m^2=m^2_c > 0$, so that 
the $m^2=0$ case lies within the broken phase.  On the other hand, the 
``renormalization-group-improved'' result obtained by resumming the 
leading-logarithmic terms \cite{cw} predicts a second-order transition 
at $m^2=0$.  The conventional view is that the latter result is trustworthy 
while the former is not.  The argument is that, for $0<m^2<m^2_c$, the 
one-loop potential's non-trivial minimum occurs only where the one-loop 
``correction'' term is as large as the tree-level term.  However, there is 
an equally strong reason to distrust the ``RG-improved'' result in the 
same region of $m^2$ and $\phi$; it amounts to re-summing a geometric series 
of leading logs that is actually a {\it divergent} series \cite{csmpla}.  
Moreover, the qualitative disagreement arises from a change in $V_{\rm eff}$ 
that in the crucial region is quantitatively tiny --- exponentially small in 
the coupling constant.  One cannot trust perturbation theory, improved or 
otherwise, at that level.  Thus, in $\LP$ theory \cite{fntescel} it is 
unsafe to draw any firm conclusion from either method; other approaches 
must be sought.

    The Gaussian approximation \cite{gauss} provides a clue.  In 3+1 dimensions 
it produces a result in agreement with the one-loop effective potential 
\cite{st}.  This is not because it contains no non-vanishing corrections beyond 
the one-loop level; it does, but those terms do not alter the functional 
form of the result.  When reparametrized in terms of a physical mass 
and field, the renormalized result is {\it exactly} the same \cite{bran}.  

    The continuum limit of $\lambda\Phi^4$ in 4 space-time dimensions is 
almost certainly `trivial' \cite{triv,book,triviality}.  Thus, a key 
consideration is what `triviality' implies about the effective potential.  
Initially, one might presume that `triviality' implies a quadratic effective 
potential, as in free-field theory.  However, that presumption accords with 
none of the approximate methods and is far from being satisfactory 
\cite{fnte1}.  Instead, we advocate the following viewpoint 
\cite{csz,csmpla,cspl}; if a theory is `trivial,' then its effective 
potential should be physically indistinguishable from the classical potential 
plus a zero-point-energy contribution of free-field form arising from 
fluctuations: 
\BE
\label{vtriv}
V_{\rm triv}(\phi)  \equiv  V_{\rm cl}(\phi) + 
\frac{1}{\vol} \sum_{\bf k} \frac{1}{2} \sqrt{{\bf k}^2 + M^2(\phi)}.
\EE
Here $M(\phi)$ denotes the mass of the shifted (`Higgs') field 
$h(x)= \Phi(x) - \phi$ in the presence of a background field $\phi$.  
After mass renormalization and subtraction of a constant term, 
$V_{\rm triv}(\phi)$ consists of $\phi^2$, $\phi^4$, and $\phi^4 \ln \phi^2$ 
terms.  Any detectable difference from this form would imply interactions 
of the $h(x)$ field --- and there are none if the theory is `trivial.'  
In other words, `triviality' implies that the exact result for the 
continuum-limit effective potential should be physically indistinguishable 
from the one-loop result.  Notice that we say ``physically indistinguishable 
from \ldots'' not ``equal to \ldots;'' it is not that multi-loop graphs 
produce no contributions but that those contributions affect both the 
effective potential and $M^2(\phi)$ in a way that preserves the functional 
form implied by (\ref{vtriv}), up to terms that vanish in the continuum limit.  

    This viewpoint explains the exact agreement between the Gaussian and 
one-loop results noted above.  Moreover, it implies that there is an infinite 
{\it class} of ``triviality-compatible'' approximations, all yielding 
the same result.  Such approximations can be arbitrarily complex provided 
they have a variational or CJT structure \cite{cjt}, with the shifted 
`Higgs' field $h(x)= \Phi(x) - \phi$ having a propagator determined 
variationally by solving a non-perturbative gap-equation.  If the approximation 
is ``triviality compatible'' then this propagator reduces to a free-field 
propagator in the infinite-cutoff limit.  In that limit all differences 
among these various approximations can be absorbed into a redefinition 
of $\lambda$, which makes no difference when the effective potential is 
expressed in terms of physical renormalized quantities \cite{bran,agodi}.  
(An explicit example of such a calculation, beyond the Gaussian approximation, 
is provided by ref \cite{rit2}.) 

    The form $V_{\rm triv}(\phi)$ differs from the prediction of 
renormalization-group-improved perturbation theory (RGIPT).  Many readers, 
we realize, will balk at any criticism of RGIPT \cite{fntergipt}.  However, 
this issue can be addressed objectively; for instance, the two predictions can 
be tested against a sufficiently precise lattice Monte-Carlo calculation 
\cite{agodi,cea}.  Data from such a lattice simulation \cite{cea} support 
our position; an excellent fit ($\chi^2/{\rm d.o.f.} \le 1$) is obtained 
with $V_{\rm triv}$, whereas the form predicted by RGIPT is unable to fit 
the data ($\chi^2/{\rm d.o.f.} \sim 6 - 10$).  This evidence certainly 
justifies us in pursuing our picture further.  

    In some respects the differences between the RGIPT and 
$V_{\rm triv}$ forms are quite small and subtle.  The following toy model 
helps to illustrate this point:  Consider a potential 
\BE
V_{\rm toy} = {{1}\over{2}} m^2 \phi^2 + 
{{\lambda}\over{4!}}\phi^4  \left( 1 + \epsilon \ln \phi^2/\mu^2 \right) ,
\EE
where $\mu$ is some mass scale and $\epsilon$ is a small parameter.  
(The real case is like $\epsilon \propto \lambda$, modulo some 
technicalities, but in this toy model we treat $\epsilon$ as a 
separate parameter.)  For $\epsilon=0$ one has a second-order phase 
transition, occurring at $m^2=0$, as one varies the $m^2$ parameter.  
However, for any positive $\epsilon$, no matter how small, one has a 
first-order transition, occurring at a positive $m^2$.  The size of $m^2$ 
involved is exponentially small in units of $\mu^2$,  
${\cal O}(\lambda \epsilon \, e^{-1/\epsilon})\mu^2$.  The vacuum value of 
$\phi^2$ is also exponentially small, ${\cal O}(e^{-1/\epsilon})\mu^2$.  
The difference between $V_{\rm toy}$ and $V_{\rm cl}$ in this region of 
$\phi$ is only ${\cal O}(\lambda \epsilon e^{-2/\epsilon})\mu^4$.  
This toy model illustrates the point that a very weak first-order phase 
transition becomes indistinguishable from a second-order transition if 
one does not look on a fine enough scale.  If one varies $m^2$ on 
a scale of $\mu^2$ one sees what looks like a second-order transition.  
Only if one varies $m^2$ on a much finer scale does one see that the 
transition is first order, exhibiting a small but non-zero jump in the 
order parameter, and occurs at a small but non-zero $m^2$.  

    We can now formulate the specific puzzle addressed in this paper.  
Suppose that spontaneous symmetry breaking does indeed coexist with a 
physical mass $m^2 \ge 0$ for the excitations of the symmetric phase.   
Those excitations would then be real particles --- as real as electrons or 
quarks (though, like quarks, they would not be directly observable); for 
brevity we call them `phions.'  The puzzle is this: How is it possible for 
the broken-symmetry vacuum --- a condensate with a non-zero density of 
phions --- to have a lower energy density than the `empty' state with no 
phions?  The $\lambda\Phi^4$ interaction corresponds to a repulsive 
`contact' interaction between phions \cite{beg} and one would think that any 
state made out of positive-mass particles with a repulsive interaction would 
necessarily have a positive energy density.  

   The solution to this puzzle is the realization that the phion-phion 
interaction is not always repulsive;  there is an induced interaction that 
is attractive.  Moreover, as $m \to 0$ the attraction becomes so long range, 
$-1/r^3$, that it generates an infrared-divergent scattering length.  
As we shall see, this long-range attraction makes it energetically 
favourable for the condensate to form spontaneously.  This leads to a simple 
picture --- a physical mechanism --- for spontaneous symmetry breaking.  

   The physics is directly related to the Bose-Einstein (BE) condensation 
of a dilute, non-ideal, Bose gas (a phenomenon observed recently in atom-trap 
experiments \cite{traps}).  The theory for this is very well established 
\cite{lhy,huang,fntebec}.  The elementary excitations of an atomic condensate 
represent not single-atom motions but collective motions --- quantized 
pressure waves, or ``phonons.''  In this language the Higgs particle is the 
phonon excitation of the phion condensate.  

    One might then ask: \ldots but how is this interesting physics consistent 
with `triviality'?  The interaction between phions should vanish in the 
infinite-cutoff limit (corresponding to shrinking the intrinsic phion size 
to zero).  How can such a `trivial' theory have a not-entirely-trivial ground 
state?  The answer is that even an {\it infinitesimal} two-body interaction 
can induce a macroscopic change of the ground state if the vacuum contains 
an {\it infinite density} of condensed phions.  Indeed we shall find 
that the condensate density is infinite in physical length units set by 
$M_h^{-1}$, the inverse Higgs mass.  Nevertheless, the condensate is 
infinitely {\it dilute} --- the density is vanishingly small on a length 
scale set by the scattering length.  This sort of subtlety reflects the 
existence of a hierarchy of scales.  One length scale, the scattering length, 
vanishes (`triviality'), while another, set by $M_h^{-1}$, remains finite.  
In fact, an intricate hierarchy of scales emerges, as we discuss in Sect. 8.  

   In what follows we use units with $\hbar=c=1$.  For simplicity we consider 
the single-component $\LP$ theory with a discrete reflection symmetry, 
$\Phi \to - \Phi$.  Since the field is Hermitian, the phion particle will 
be its own antiparticle.  In Sect. 2 we discuss the interparticle potential 
between phions.  Then in Sect. 3 we estimate the energy density of a phion 
condensate, with a given particle density, in a very simple and intuitive 
way.  The result is confirmed in Sect. 4 by a calculation based on the 
Lee-Huang-Yang (LHY) treatment of a non-relativistic Bose gas.  The resulting 
energy-density expression, equivalent to the field-theoretic effective 
potential, yields a phase transition which we analyze in Sect. 5.  The 
excitations of the condensate are `phonons' (Higgs particles) and in principle 
the physics can be described either in terms of phions or in terms of phonons.  
Sect. 6 discusses how the ``renormalized field'' associated with phonons is 
related to the original (``bare'') field associated with phions.  In Sect. 7 
the effective potential is written in manifestly finite form in terms of the 
renormalized field.  Sect. 8 provides a brief summary and discusses the 
intricate hierarchy of length scales that arise.  We conclude with some 
speculations about the possible wider implications of these ideas.

\setcounter{equation}{0}
\section{Interparticle potential between phions}  

     In QED there is a well-known equivalence between the photon-exchange 
Feynman diagram and the Coulomb potential $1/r$.  Similarly, pion exchange 
gives rise to the Yukawa potential ${\rm e}^{-mr} \! /r$ in nuclear physics.  
The exchange of two massless neutrinos gives rise to a long-range $1/r^5$ 
potential \cite{feinberg2}.  In $\LP$ theory it is well known that the 
fundamental interaction vertex corresponds to a $\dl$ interaction \cite{beg}.  
However, as we now explain, the exchange of two virtual phions gives rise 
to an attractive long-range $-1/r^3$ interaction.  

      Consider the elastic collision of two particles of mass $m$ in the 
centre-of-mass frame.  Let ${\bf q}$ denote the 3-momentum transfer; let 
$\theta$ be the scattering angle; and let $E = \sqrt{{\bf p}^2 + m^2}$ be 
the energy of each particle.  A scattering matrix element ${\cal M}$, 
obtained from Feynman diagrams, can be associated with an `equivalent 
interparticle potential' that is is basically the 3-dimensional Fourier 
transform of ${\cal M}$:
\BE
\label{veq}
V({\bf r}) = \frac{1}{4 E^2} \int \! \frac{d^3 q}{(2 \pi)^3} \,
{\rm e}^{i {\bf q}.{\bf r}} \, {\cal M}.
\EE 
(For a detailed discussion see the review article of Feinberg {\it et al} 
\cite{feinberg}.)  This `equivalent potential' is a function of the relative 
position ${\bf r}$ (conjugate to ${\bf q}$).  In general it also depends 
parametrically on the energy, $E$, though this complication disappears in the 
non-relativistic limit, where $E \sim m$.  

   For the $\lambda \Phi^4$ theory the lowest-order Feynman diagram 
(see Fig. 1) gives ${\cal M}_o = \lambda$ and the resulting potential is
\BE
\label{vbare}
V_o({\bf r}) = \frac{1}{4 E^2} \lambda \dl. 
\EE
In a non-relativistic treatment of the theory this is the only interaction.  
The `triviality' property is then reflected in the well-known fact that in 
quantum mechanics a 3-dimensional $\delta$-function interaction gives zero 
scattering amplitude \cite{beg}.  

     Relativistically there are additional contributions to the equivalent 
interparticle potential, notably those produced by the one-loop
`fish' diagrams (see Fig. 2) corresponding to the three Mandelstam variables 
$s,t,u$ ($s=4E^2$, $t=-{\bf q}^2$, $s+t+u=4m^2$).  To evaluate these 
contributions, we first note that if ${\cal M}$ depends only on 
$q \equiv \mid\!{\bf q}\!\mid $, then Eq.~(\ref{veq}) reduces to 
\BE
V(r) = \frac{1}{4 E^2} \frac{1}{(2 \pi)^2} \int_0^\infty q^2 dq 
\frac{2 \sin qr}{q r} {\cal M}(q)
\EE
\BE 
\label{veq2}
\quad \quad \quad = \frac{1}{8 \pi^2 E^2} \, \frac{1}{r^3} 
\int_0^\infty dy \, y \sin y \, {\cal M}(q=y/r).
\EE
This already shows that $V(r)$ is spherically symmetric and naturally has 
a factor $1/r^3$.  

     To evaluate the contribution from $t$-channel scattering, we begin with 
the case $m=0$, where the matrix element is simple:
\BE
\label{mtexch}
{\cal M}_{t-{\rm exch}}(q) = \frac{\lambda^2}{16 \pi^2} \ln(q/\Lambda),
\EE
where $\Lambda$ is the ultraviolet cutoff.  Substituting into Eq.~(\ref{veq2}) 
we find an integral that is not properly convergent but which can be 
made convergent by including a factor ${\rm e}^{-\epsilon y}$ and then 
taking the limit $\epsilon \to 0$ (physically, this corresponds to 
smearing out the point vertices).  In this sense we have \cite{gr}
\BE 
\label{gr1}
\int _0^\infty dy \, y \sin y =0, 
\EE 
and  
\BE 
\label{gr2}
\int _0^\infty dy \, y \ln y \, \sin y = - \frac{\pi}{2}.
\EE 
The first equation implies that those terms independent of $q$ in the 
matrix element do not give contributions to the potential for values of 
$r\neq 0$.  Such terms, however, bring a contribution of the type $\dl$, 
as we see by returning to the form (\ref{veq}).  The $s$-channel 
amplitude, for example, gives only a contribution of this type.  These 
delta-function contributions can be absorbed into a redefinition of $\lambda$, 
the strength of the repulsive potential in Eq.~(\ref{vbare}).  In this way, 
we can include all possible diagrammatic contributions to the short-range 
repulsive interaction.  Then, $\lambda$ would become an effective parameter 
representing the actual physical strength of the repulsive contact 
interaction, rather than the bare coupling entering in the Lagrangian 
density.  

     Substituting Eq. (\ref{mtexch}) into (\ref{veq2}) we find an 
attractive, long-range potential:
\BE
\label{pot}
V_{t-{\rm exch}}(r) = - \frac{\lambda^2}{256 \pi^3 E^2} \frac{1}{r^3} .
\EE
An equal contribution is obtained from the $u$-channel diagram; in QM terms 
it corresponds to the amplitude $f(\pi -\theta)$ that must be added to 
$f(\theta)$ when dealing with identical-particle scattering.  Note that, 
as a consquence of Eq.~(\ref{gr1}), there is no dependence on $\Lambda$ 
in the $-1/r^3$ potential.  

    Taking into account the mass $m$ of the exchanged particles yields 
the result (\ref{pot}) multiplied by a factor $2mr K_1(2mr)$, where $K_1$ 
is the modified Bessel function of order unity.  This factor tends to unity 
as $mr \to 0$ and for large values of $mr$ tends to 
$\sqrt{\pi m r} \, {\rm e}^{-2 m r}$.  The exponential factor is like that 
of the Yukawa potential except that, since there are two exchanged particles, 
it is ${\rm e}^{-2 m r}$ instead of ${\rm e}^{- m r}$.  
Physically, the $-1/r^3$ potential arises from two short-range repulsive 
interactions linked by the quasi-free propagations of two virtual particles.  
(Exchanges of more than 2 particles over macroscopic distances would lead to 
contributions with a faster power-law fall-off.)  Thus, higher-order 
contributions are accounted for by the same redefinition of $\lambda$ 
mentioned above.  If the short-range repulsive interaction has an actual 
strength $\lambda$, then the $-1/r^3$ attractive interaction is proportional 
to $\lambda^2$.  

      In summary: the interparticle potential is essentially given by 
the sum of a repulsive core, $\dl$, and an attractive term $-1/r^3$ that 
is eventually cut off exponentially at distances greater than $1/(2m)$.  
The long-range attraction between the phions has important effects, as we 
shall see in the next section.  

\setcounter{equation}{0}
\section{Condensate energy density: a simple estimate } 

    Consider a large number $\num$ of phions contained in a large box of 
volume $\vol$.  As in statistical mechanics, the thermodynamic limit requires 
$\num \to \infty$ and $\vol \to \infty$ with the density $n \equiv \num/\vol$ 
being fixed.  The `empty' state corresponds to the special case $n=0$.  
Since phions can be created and destroyed, the equilibrium value of $n$ is 
to be determined by minimizing the energy density in the box.  In this 
section we estimate the ground-state energy density for a given $n$ in a 
very simple and intuitive way.  Some tedious subtleties affecting numerical 
factors are ignored here.  A proper calculation will be provided in the next 
section.  

     Assuming the density $n$ is low, the relevant contributions to the 
total energy of the system are just the rest-masses $\num m$ and the 
two-body interaction energies.  Effects from three-body or multi-body 
interactions will be negligible provided the gas of phions is dilute.  
The two-body contribution is the number of pairs ($\half \num (\num-1) \approx 
\half \num^2$) multiplied by the average potential energy between a pair of 
phions: 
\BE
\label{media}
\bar{u} \sim {{1}\over{\vol}}\int \! d^3r \, V(r) .
\EE
This averaging assumes that the particles are uniformly distributed over the 
box, which is valid since at zero temperature almost all the particles are 
condensed in the ${\bf k}=0$ mode.  Thus, the total energy of the ground 
state is 
\BE
E_{\rm tot} = \num m + \half \num^2 \bar{u},   
\EE
yielding an energy density 
\BE
{\cal E}  \equiv E_{\rm tot} /\vol = 
n m +  \half n^2 \int \! d^3 r \, V(r).  
\EE

    The potential $V(r)$ consists of the $\dl$ term (\ref{vbare}) and 
the $-1/r^3$ term (\ref{pot}) (times 2 to include the $u$-channel).  
We may set $E=m$ since almost all phions have ${\bf k} = 0$.  Thus, 
we find 
\BE
\label{estrr}
{\cal E}
= nm + {{\lambda n^2}\over{8m^2}} -
{{\lambda^2 n^2}\over{64\pi^2 m^2}} \int \! {{dr}\over{r}} .
\EE
The integral over $r$ can be cut off at small $r$ by introducing a 
`hard-core radius' $r_o$, corresponding to an ultraviolet regularization 
that smears out the $\dl$ point-like interaction.  In addition, 
to avoid an infrared divergence, the integral must also have some 
large-distance cutoff, $r_{\rm max}$.  Since, as noted in the last section, 
the phion mass introduces an exponential factor $ {\rm e}^{-2 m r}$ into 
the potential, we have $r_{\rm max} \le 1/(2 m)$.  However, if $m$ is very 
small another consideration is actually more important; namely, that the 
long-distance attraction between two phions becomes ``screened'' by other 
phions that interpose themselves.  This immediately implies an $r_{\rm max}$ 
that depends on the density $n$.  In fact, $r_{\rm max}$ is naturally given by 
$1/(2 M)$, where $M$ is the mass of the quasiparticle excitations of the 
condensate, and it is easily seen that $M^2$ is proportional to $n$ when 
$m$ is small \cite{fnteM}.  

     Hence, ${\cal E}$ is given by a sum of $n$, $n^2$ and $n^2 \ln n$ terms 
which represent, respectively, the rest-mass energy cost, the repulsion 
energy cost, and the energy gain from the long-range attraction.  If the 
rest-mass $m$ is small enough, then the $n^2 \ln n$ term's negative 
contribution can result in an energy density whose global minimum is not 
at $n=0$ but at some specific, non-zero density $n_v$.  That is, even though 
the `empty' state is locally stable, it can decay by spontaneously generating 
particles so as to fill the box with a dilute condensate of density $n_v$.  

    The result can be translated into field-theory terms since the particle 
density $n$ is proportional to the intensity of the field, $\phi^2$.  
In fact, as shown in the next section, one has $n = \half m \phi^2$.  
The energy density as a function of $n$ then becomes the field-theoretic 
{\it effective potential}: ${\cal E} (n)\equiv V_{\rm eff}(\phi)$.  
(Of course this ``potential'' for the {\it field} is not to be confused with 
the ``potential'' between particles, $V(r)$.)  The estimate above leads to 
\BE
\label{density}
V_{\rm eff} (\phi)= {{1}\over{2}}m^2\phi^2 + {{\lambda\phi^4}\over{32}}
-{{\lambda^2\phi^4}\over{256\pi^2}}
\ln {{r_{\rm max}(\phi) }\over{r_o}}.  
\EE
We can identify $r_o$ with the reciprocal of an ultraviolet cutoff $\Lambda$ 
and $r_{\rm max}(\phi)$ with $1/(2M)$, where $M^2 \propto n \propto \phi^2$.  
Thus, the essential form of the result is here.  (The incorrect numerical 
factors could be straightened out with enough care, we believe.)  

    This simple approach gives some important insight.  The reason there 
are no $n^3, n^4 \ldots$ ($\phi^6, \phi^8, \ldots$) terms is the diluteness 
of the gas.  Furthermore, if the attractive potential had fallen off faster 
than $1/r^3$ then ${\cal E}$ would have had only $n$ and $n^2$ terms.  
The crucial $n^2 \ln n$ term arises from the infrared divergence of the 
integral in (\ref{estrr}) which is tamed only by the screening effect of 
the background density $n$.  Thus the $\phi^4 \ln \phi^2$ term has two 
complementary interpretations:  In field language it arises from the 
zero-point energy of the field fluctuations, while in particle language it 
arises from the long-range attraction between phions.  

\setcounter{equation}{0}
\section{Condensate energy density: calculation \`{a} la LHY }

\par In this section we compute the energy density using a relativistic 
version of the original Lee-Huang-Yang (LHY) analysis of Bose-Einstein 
condensation of a non-ideal gas \cite{lhy,huang}.  We emphasize that their 
analysis invokes neither a weak-coupling nor a semiclassical approximation.  
They appeal to two approximations: (1) low energy, so that the scattering 
is pure $s$-wave and (2) low density (diluteness).  In terms of the 
phion-phion scattering length 
\BE
a = \frac{\lambda}{8 \pi E}  
\EE
these conditions are:
\BE
\label{low}
 {\mbox{\rm `low-energy':}} \, \, \quad k a \ll 1,
\EE
\BE
\label{lowd}
 {\mbox{\rm `diluteness':}} \, \, \quad na^3 \ll 1.
\EE
Note that `low energy' in the above sense does not imply `non-relativistic' 
--- because here (quite unlike the situation in atomic physics) we may have 
$m \ll 1/a$.  

    We start from the $\LP$ Hamiltonian: 
\BE 
\label{hamiltonian}
H = \, : \! \int \! d^3 x \left[ \frac{1}{2} \left( \Pi^2 + 
(\del \Phi)^2 + m^2 \Phi^2 \right) + 
\frac{\lambda}{4!} \Phi^4 \right] \! : \; . 
\EE
The system is assumed to be contained within a finite box of volume $\vol$ 
with periodic boundary conditions.  There is then a discrete set of allowed 
modes ${\bf k}$.  In the end we will take the infinite-volume limit and the 
summation over allowed modes will go over to an integration: 
$\sum_{\bf k} \to \vol \int d^3k/(2 \pi)^3$.  

    Annihilation and creation operators, $a_{\bf k}$, $a^{\dagger}_{\bf k}$, 
are introduced through the plane-wave expansion
\BE
\label{phime}
\Phi({\bf x},t) = 
\sum_{\bf k} \frac{1}{\sqrt{2 \vol E_k}} 
\left[ a_{\bf k} {\rm e}^{i {\bf k}.{\bf x}} + 
a^{\dagger}_{\bf k} {\rm e}^{-i {\bf k}.{\bf x}} 
\right] ,
\EE
where $E_k=\sqrt{{\bf k}^2 + m^2}$.  The $a_{\bf k}$'s are time dependent 
(in the free-field case they would be proportional to ${\rm e}^{-i E_k t}$) 
and satisfy the commutation relations 
\BE
[ a_{\bf k}, a^{\dagger}_{{\bf k}'} ] = \delta_{{\bf k},{\bf k}'}.
\EE
The Hamiltonian includes ``normal ordering'' symbols :$ \ldots \,$: so that 
so that the quadratic part of $H$ is just \cite{massren}
\BE
\label{quadratic}
H_2 = \sum_{\bf k} E_k a^{\dagger}_{\bf k} a_{\bf k}.
\EE

     For comparison with the non-relativistic calculation, it is convenient 
to subtract from the Hamiltonian a term $\mu_c \hat{\num}$, where 
$\hat{\num}$ is the operator that counts the number of phions: 
\BE
\label{nop}
\hat{\num} = \sum_{\bf k}  a^{\dagger}_{\bf k} a_{\bf k} .
\EE
At the end we shall set the chemical potential $\mu_c=0$.  However, in the 
non-relativistic context one should put $\mu_c =m$ to take into account that, 
then, the rest-mass energy is not counted as part of a particle's kinetic 
energy.  Therefore, the correct definition of the total energy of the system, 
in the non-relativistic context, is obtained by subtracting $m$ for each 
particle, so that $H_{\rm NR} = H - m \hat{\num}$.  
     
    When a system of $\num$ bosons undergoes Bose-Einstein condensation, then 
the lowest energy mode becomes macroscopically populated, below some critical 
temperature.  That is, there are $\num_0$ particles in the ${\bf k}=0$ 
mode, with $\num_0$ being a finite fraction of the total number $\num$.  
At zero temperature, if the gas is dilute, almost all the particles are in 
the condensate; $\num_0{\small (T=0)} \sim \num$.  In fact, the fraction not 
in the condensate is of order $\sqrt{n a^3}$ \cite{lhy,huang} and so is 
negligible in the dilute approximation.  We then have 
$a^{\dagger}_0 a_0 \sim \num$, and so we can consider $a_0$ to be essentially 
the c-number, $\sqrt{\num}$. (Of course $a_0$ still has an operator part, 
but any relevant matrix elements of this part are only of order unity, 
negligible in comparison to the c-number part $\sqrt{\num}$.)  From the 
expansion (\ref{phime}) we then get the expectation value 
\BE
\label{phn}
\phi = \langle \Phi \rangle = \frac{1}{\sqrt{2 \vol m}} ( a_0 + a_0^\dagger ) 
= \sqrt{\frac{2\num}{\vol m}}.  
\EE
Hence, the particle density $n \equiv \num/\vol$ is given by 
\BE
\label{neq}
n = \half m \phi^2,
\EE
as anticipated in the previous section.  With this identification, setting 
$a_0 = a_0^\dagger = \sqrt{\num}$ is equivalent to shifting the quantum 
field $\Phi$ by a constant term $\phi$.  

     Making this substitution yields 
\BE
\label{heffe}
H_{\rm eff} - \mu_c \hat{\num} = \vol \left[ (m-\mu_c)n + 
\frac{\lambda n^2}{6 m^2} \right] 
+ \sum_{{\bf k} \neq 0} \left[ 
a^\dagger_{\bf k} a_{\bf k} \left( E_k - \mu_c + \frac{\chi}{2 E_k} \right) + 
\frac{\chi}{4 E_k} \left( a_{\bf k} a_{-{\bf k}} + 
a^\dagger_{\bf k} a^\dagger_{-{\bf k}} \right) \right]
\EE
with 
\BE
\chi = \frac{\lambda n}{m} = \half \lambda \phi^2 .
\EE
As stressed in Ref. \cite{aaa}, this result contains all interactions of 
condensate particles between themselves and all interactions between 
condensate and non-condensate particles.  It neglects interactions among the 
non-condensate particles, which is justified because there are so few of them; 
their density is smaller than $n$ by a factor of $\sqrt{na^3}$ \cite{huang}.  
We stress that the justification here is not weakness of interaction 
but scarcity of interactors; i.e., {\it low density} and not {\it weak 
coupling}.  

     To diagonalize the Hamiltonian (\ref{heffe}) we can define new 
annihilation and creation operators $b_{\bf k}, b^\dagger_{\bf k}$ (for 
${\bf k} \neq 0$).  The linear transformation 
\BE
\label{bogo}
a_{\bf k} = \frac{1}{\sqrt{1-\alpha_k^2}} 
\left( b_{\bf k} - \alpha_k b_{- {\bf k}}^{\dagger} \right) 
\EE
and its Hermitian conjugate are called the Bogoliubov transformation.  
The quanta annihilated and created by the operators 
$b_{\bf k}, b^\dagger_{\bf k}$ are called `phonons' or `quasiparticles' 
to distinguish them from the `particles' associated with the original operators 
$a_{\bf k}, a^\dagger_{\bf k}$.  The function $\alpha_k$ is fixed by the 
requirement that in the Hamiltonian Eq.~(\ref{heffe}), the coefficients of 
the $b_{\bf k}b_{-\bf k}$ and $b_{\bf k}^{\dagger}b_{-\bf k}^{\dagger}$ 
terms vanish.  This fixes 
\BE 
\label{alpharel}
\alpha_k = 1+x^2-x\sqrt{x^2+2},  \quad \quad 
x^2 \equiv \frac{2}{\chi} E_k (E_k - \mu_c) .
\EE
The result then takes the form 
\BE
\label{heffrelb}
H_{\rm eff} - \mu_c \hat{\num} = E_{\rm tot} + 
\sum_{{\bf k} \neq 0} \widetilde{E_k} b^\dagger_{\bf k} b_{\bf k}.
\EE
Apart from the constant term $E_{\rm tot}$, which we discuss below, this 
is analogous to Eq.~(\ref{quadratic}) but with a different spectrum 
(energy-momentum relationship) for the quasiparticles:
\BE 
\label{etk}
\widetilde{E_k} = \left( \frac{1+\alpha_k}{1-\alpha_k} \right) E_k ,
\EE
which yields 
\BE
\widetilde{E_k} = (E_k - \mu_c) \sqrt{ 1 + \frac{\chi}{E_k(E_k - \mu_c)} } .
\EE
In the non-relativistic limit, setting $\mu_c=m$ and 
$E_k=\sqrt{ {\bf k}^2 +m^2} \approx m + k^2/(2m) + \ldots$, one obtains the 
famous `Bogoliubov spectrum':
\BE
\widetilde{E_k}^{\rm NR} = {{ k}\over{2m}}
 \sqrt{ {\bf k^2} + 2 \chi },
\EE
with its characteristic linear behaviour, as ${\bf k} \to 0$, for the 
`phonon' excitations of a dilute Bose gas at low temperature.  In the 
relativistic case, where $\mu_c=0$, one has instead 
\BE
\label{enew}
\widetilde{E_k} = \sqrt{E_k^2 + \chi} = \sqrt{{\bf k}^2 + m^2 + \chi}.
\EE 
This has the normal form for a relativistic energy-momentum relation 
and we can identify the mass of the `quasiparticle' excitations as:
\BE
\label{msqnew}
M^2(\phi) = m^2 + \chi = m^2 + \frac{1}{2} \lambda \phi^2.
\EE

     The constant term in Eq. (\ref{heffrelb}) above is given (for $\mu_c=0$) 
by
\BE
\label{egsnew}
E_{\rm tot} =
\vol \left[ nm + \frac{\lambda n^2}{6 m^2} \right] - 
\frac{\chi}{4} \sum_{k \neq 0} \frac{\alpha_k}{E_k}.  
\EE
The term in square brackets is just the `classical' energy density
\BE
\label{classico}
V_{\rm cl} (\phi) = \frac{1}{2}m^2 \phi^2 + \frac{\lambda}{4!} \phi^4 
\EE
as sees by substituting $n=\half m \phi^2$.  The last term in (\ref{egsnew}) 
arises because 
$b_{\bf k} b^{\dagger}_{\bf k} = b^{\dagger}_{\bf k} b_{\bf k} + 1$.  
To evaluate it we substitute for $\alpha_k$ from Eq. (\ref{alpharel}), 
and use $x\sqrt{x^2 + 2} = 4 E_k \widetilde{E_k}/(\lambda \phi^2)$ to obtain 
\BE
\label{veehehe}
V_{\rm eff} = \frac{1}{2}m^2 \phi^2 + \frac{\lambda}{4!} \phi^4 
+ I_1(M) - I_1(m) - \frac{1}{4}\lambda \phi^2 I_0(m),
\EE
where 
\BE 
I_1(M) \equiv \int \! \frac{d^3 k}{(2 \pi)^3} \frac{1}{2} \widetilde{E_k}, 
\quad \quad 
I_1(m) \equiv \int \! \frac{d^3 k}{(2 \pi)^3} \frac{1}{2} E_k, 
\quad \quad 
I_0(m) \equiv \int \! \frac{d^3 k}{(2 \pi)^3} \frac{1}{2 E_k}. 
\EE
In field-theory language $I_1(M)$ represents the zero-point fluctuations 
of a free scalar field of mass $M=M(\phi)$, and the last two terms of 
Eq.(\ref{veehehe}) represent the subtractions associated with the 
normal ordering of the Hamiltonian (\ref{hamiltonian}).  Such 
subtractions remove the quartic divergence $\sim \Lambda^4$ and 
the quadratic divergence $\sim \Lambda^2$ that are contained in $I_1(M)$, 
leaving only a logarithmic divergence $\ln \Lambda$.  Explicit
calculation gives 
\BE
\label{final}
V_{\rm eff}(\phi) = \frac{1}{2}m^2 \phi^2 + \frac{\lambda}{4!} \phi^4 
+\frac{\lambda^2}{256 \pi^2} \phi^4 
\left[ \ln (\half \lambda \phi^2 /\Lambda^2 ) - \frac{1}{2} 
+F\left(\frac{m^2}{\half \lambda \phi^2}\right)
\right] ,
\EE 
where 
\BE
F(y) = \ln(1+y) + \frac{y(4+3y)}{2(1+y)^2}, \quad \quad \quad 
y \equiv  \frac{m^2}{\half \lambda \phi^2}.  
\EE

   The result (\ref{final}) coincides with the famous one-loop result 
\cite{cw}.  However, our point is that the result is justified by the 
`low-energy' and `diluteness' assumptions, without appealing to perturbative 
or semiclassical approximations.  This point is well known in the 
non-relativistic case \cite{huang,aaa}.  For small $m$, when the $F(y)$ 
term can be neglected, the result (\ref{final}) has the same structure 
found in the intuitive calculation in the preceding section, Eq. 
(\ref{density}).

\setcounter{equation}{0}
\section{The phase transition }

    We now analyze the energy-density expression Eq. (\ref{final}) to find 
where the phase transition occurs.  It is easy to guess that $m^2$ will have 
to be small, so we expect $y \equiv m^2/(\half \lambda \phi^2) \ll 1$ 
everywhere except very near the origin.  This implies that the phonon mass 
$M^2(\phi)$, Eq. (\ref{msqnew}), will be much larger than $m^2$ and will 
become essentially $\half \lambda \phi^2$.  In this regime the mass will be  
relevant only in the rest-mass energy term ${{1}\over{2}}m^2\phi^2$ 
and we may neglect the $F(y)$ term.  We proceed to do so, but we shall return 
at the end to verify that this is justified.  Thus, we start from 
\BE
\label{veffeq}
V_{\rm eff} = \frac{1}{2} m^2 \phi^2_B + \frac{\lambda}{4!} \phi^4_B 
+ \frac{\lambda^2}{256 \pi^2} \phi^4_B 
\left[ \ln (\half \lambda \phi^2_B /\Lambda^2 ) - \frac{1}{2} \right] ,
\EE
where we have added a `$B$' subscript to $\phi$ to emphasize that it is 
a `bare' (unrenormalized) field.  
If $V_{\rm eff}$ has a pair of extrema $\phi_B = \pm v_B$, then 
$v_B$ is a solution of $dV_{\rm eff}/d \phi_B =0$, which gives 
\BE
\label{vbeqa}
m^2 + \frac{\lambda}{6} v_B^2 + 
\frac{\lambda^2}{64 \pi^2} v_B^2 
 \ln ( \half \lambda v_B^2/\Lambda^2)  =0.
\EE
This condition allows us to eliminate $\Lambda$ in favour of $v_B$, so that 
the effective potential (\ref{veffeq}) can be expressed equivalently as 
\BE
\label{equiv}
V_{\rm eff} = \frac{1}{2} m^2 \phi^2_B (1- {{\phi^2_B}\over{2v^2_B}})
+ \frac{\lambda^2}{256 \pi^2} \phi^4_B 
\left[ \ln {{\phi^2_B}\over{v^2_B}}  - \frac{1}{2} \right] .
\EE
We denote by $v_0$ the value of $v_B$ in the case $m^2 = 0$; it is given by 
\BE
\label{v0eq}
v_0^2 = \frac{2 \Lambda^2}{\lambda}  
\exp \left( - \frac{32 \pi^2}{3 \lambda}  \right).
\EE
The original equation (\ref{vbeqa}) can then be re-written as:
\BE
\label{vbeq}
f(v_B^2) \equiv - \frac{\lambda^2}{64 \pi^2} v_B^2 
\ln (v_B^2/v_0^2) = m^2.
\EE
A graph of $f(v_B^2)$ starts from zero at $v_B^2=0$, reaches a maximum 
at $v_B^2={\rm e}^{-1}v_0^2$ and then decreases, becoming negative when 
$v_B^2 > v_0^2$.  Equating this to $m^2$ we see that: (i) if  $m^2$ is 
positive and larger than the maximum value of $f$ then Eq. (\ref{vbeq}) 
has no real roots.  In this case $V_{\rm eff}$ has a single minimum located 
at $\phi_B=0$.  (ii) if $m^2$ is positive but not too large, then 
Eq. (\ref{vbeq}) has two roots.  In that case $V_{\rm eff}$ has a local 
minimum at $\phi_B=0$, then a maximum (at the smaller $v_B$ root) and then a 
minimum (at the larger root, the true $v_B$).  (iii) if $m^2$  is negative 
there is a unique root, with $v^2_B$ greater than $v^2_0$.  In that case, 
the origin $\phi_B=0$ is a maximum of $V_{\rm eff}$ and $v_B$ is an absolute 
minimum.  

    Case (i) is not very interesting since it does not show condensation 
and spontaneous symmetry breaking.  Case (iii) shows spontaneous symmetry 
breaking but, with $m^2$ negative, the phions would not be particles 
in the ordinary sense.  Our interest in this paper is with case (ii) 
where $m^2$ is positive but less than the maximum of the function $f$:
\BE
m^2 <  \frac{ \lambda^2}{64 \pi^2} {{v_0^2}\over{ {\rm e} }}.
\EE
This condition ensures that non-trivial minima of $V_{\rm eff}$ exist.  
A stronger bound on $m^2$ is needed if the minimum at $v_B$ is 
to have lower energy density than the symmetric vacuum.  For this we need 
\BE
\label{vatvb}
V_{\rm eff}(\phi_B=\pm v_B)={{1}\over{4}}
(m^2- \frac{ \lambda^2}{128 \pi^2} v_B^2 )v^2_B 
\EE
to be negative.  Combined with Eq. (\ref{vbeq}), this gives 
\BE
\label{trans}
m^2\leq 
\frac{ \lambda^2}{128 \pi^2} {{v_0^2}\over{ \sqrt{ {\rm e} } }}
\equiv m^2_c 
\EE
as the condition for condensation to be energetically favoured.  
At $m^2=m^2_c$ the symmetric phase at $v_B=0$ and the condensate phase at 
$v^2_B=v^2_0/\sqrt {{\rm e}}$ have equal energy density, allowing the 
possibility of co-existence of the two phases.

    The Higgs-boson mass $M_h$ corresponds to $M(\phi)$ in the physical 
vacuum, $\phi_B=v_B$.  Thus, from Eq. (\ref{msqnew}) we have 
\BE
\label{mheq}
M_h^2 \equiv M^2(\phi_B = v_B) = m^2 +  \half \lambda v^2_B .
\EE
We shall neglect the $m^2$ term and justify this later.  Noting that 
$v_B^2$ lies between ${\rm e}^{-1}v_0^2$ and $v_0^2$, we see from 
(\ref{v0eq}) that 
\BE
\label{mheqb}
M_h^2 = {\cal O} \left( \Lambda^2 
\exp\left( -\frac{32\pi^2}{3\lambda} \right) \right).
\EE
We want in the end to take the cutoff $\Lambda$ to infinity, but such 
that the essential physics remains independent of $\Lambda$.  In this 
case the crucial condition is that the physical value of the Higgs mass, 
$M_h^2$, should remain finite.  The only way to obtain that result is to take 
$\lambda$ to zero as $\Lambda \to \infty$.  Indeed, re-arranging 
(\ref{mheqb}) we see that $\lambda$ must behave as: 
\BE 
\frac{\lambda}{16 \pi^2} \sim \frac{2}{3}
\frac{1}{\ln(\Lambda^2/M_h^2)}.
\EE
Thus, the coupling constant $\lambda$ must depend on the cutoff, and, in 
particular, must tend to zero like $1/\ln \Lambda$.  In that same limit 
$v_B^2$ must diverge as $\ln \Lambda$ so that the product $\lambda v_B^2$, 
and hence $M_h^2$, is finite \cite{book2}.  We also see now, from 
(\ref{trans}), that $m^2 \le m_c^2 \sim \lambda^2 v_B^2$, so that 
$m^2 = {\cal O}(\lambda M_h^2)$.  Since $\lambda \to 0$, it follows that 
$m^2$ becomes vanishingly small, while $M_h^2$ remains finite.  We were 
therefore justified in neglecting the $m^2$ term in Eq. (\ref{mheq}).  
In summary, we have (for $\Lambda$ in units of $M_h$) 
\BE
\label{gb1}
\lambda = {\cal O}(1/\ln \Lambda), \quad \quad 
m^2 = {\cal O}(1/\ln \Lambda), \quad \quad v_B^2 = 
{\cal O}(\ln \Lambda).  
\EE
This type of behaviour ensures that the Higgs mass and the energy density 
at the minimum (\ref{vatvb}) are finite and so we have a physically 
significant limit when $\Lambda \to \infty$.  If $\lambda$ were larger, 
then $M_h$ would go to infinity (in particular, $M_h= {\cal O}(\Lambda)$ 
for any small but finite value of $\lambda$).  If $m^2$ were larger, then 
the term $\half m^2 \phi^2_B$ would dominate $V_{\rm eff}$ which would have 
only a minimum at $\phi_B=0$.  The interesting region is close 
to the phase transition where the no-phion state at $\phi_B=0$ and 
the condensate state at $\phi_B=v_B$ are very close in energy.  In that 
region the elementary excitations of both vacua, phions and Higgs bosons, 
have vastly different masses as anticipated in the introduction.  

     Using the result (\ref{gb1}) it is now straightforward to justify 
the neglect of the $F(y)$ term in (\ref{final}), at least for $\phi_B$'s 
that are comparable to $v_B$, where it becomes ${\cal O}(1/\ln \Lambda)$.  
At much smaller $\phi_B$ the $F(y)$ term does play some role; it serves 
to smooth out what would otherwise be a singularity in the fourth derivative 
at the origin.  

     $V_{\rm eff}$ has an important qualitative difference from the 
classical potential (\ref{clpot}).  The latter has a double-well form 
only for negative $m^2$ values and has a phase transition of second 
order at the value $m^2=0$.  With $V_{\rm eff}$ the phase transition occurs 
at $m^2=m^2_c$, Eq. (\ref{trans}), and the `order parameter' 
$\langle \Phi \rangle$ jumps from zero to ${\rm e}^{-1/4} v_0$.  
This first-order character, with the possibility of phase coexistence, 
remains in the limit $\Lambda \to \infty$, despite the fact that 
$m^2_c \to 0$ in units of $M_h^2$.  

\setcounter{equation}{0}
\section{Phions and phonons }

The effective potential in terms of the bare field $\phi_B$ is an extremely 
flat function because the $\phi_B=0$ vacuum and the $\phi_B=\pm v_B$ vacua are
infinitely far apart ($v^2_B \sim \ln \Lambda$) but their energy densities 
differ only by a finite amount.  To plot a graph of $V_{\rm eff}(\phi)$ 
one would naturally want to re-define the scale of the horizontal axis by 
defining a `renormalized' or `re-scaled' field $\phi_R$.  However, this 
does not correspond to a traditional wavefunction renormalization.  Instead, 
it requires the following procedure \cite{csz,cspl}: (i) Decompose the 
full field $\Phi_B(x)$ into its zero- and finite-momentum pieces:
\BE
\Phi_B(x) = \phi_B + h(x),
\EE
where $\int \! d^4x \, h(x) =0$.  (ii) Re-scale the zero-momentum part 
of the field: 
\BE
\phi_B^2 = Z_{\phi} \phi_R^2 
\EE
with a $Z_{\phi}$ that is {\it large}, of order $\ln \Lambda$. (The condition 
determining $Z_{\phi}$ is discussed below).  (iii) The finite-momentum 
modes $h(x)$ remain unaffected.  

    This procedure is very natural in terms of the LHY calculation in Sect. 4, 
where singling out the ${\bf k}=0$ mode is crucial.  From Eq. (\ref{phime}), 
excluding ${\bf k}=0$, we have 
\BE
h({\bf x},t)  \equiv  \sum_{{\bf k} \neq 0} \frac{1}{\sqrt{2 \vol E_k}} 
\left[ a_{\bf k} {\rm e}^{i {\bf k}.{\bf x}} + 
a^{\dagger}_{\bf k} {\rm e}^{-i {\bf k}.{\bf x}} 
\right] .  
\EE
Substituting the Bogoliubov transformation (\ref{bogo}), re-organizing the 
summation using ${\bf k} \to -{\bf k}$ in some terms, and finally using 
(\ref{etk}) we see that
\BE
h({\bf x},t) = \sum_{{\bf k} \neq 0} \frac{1}{\sqrt{2 \vol \widetilde{E_k}}} 
\left[ b_{\bf k} {\rm e}^{i {\bf k}.{\bf x}} + 
b^{\dagger}_{\bf k} {\rm e}^{-i {\bf k}.{\bf x}} 
\right] .
\EE
Thus, the finite-momentum part of the field is not re-scaled; it takes the 
canonical form in terms of phion or phonon variables.  However, note that 
the Bogoliubov transformation (\ref{bogo}) applies only to the 
${\bf k} \neq 0$ modes; ``$b_0, b_0^\dagger$'' remain undefined and are 
not necessarily related to $a_0, a_0^\dagger$ ($\sim \sqrt{\num}$) in the 
same way.  Indeed, the $a_{\bf k}$'s are discontinuous as ${\bf k} \to 0$ 
because $\num$ phions occupy the ${\bf k}=0$ mode.  However, for phonons 
the physics is continuous as ${\bf k} \to 0$; that condition will fix 
$Z_\phi$, as we show below.  
   
    In field-theory language the corresponding discussion is as follows.  
For the finite-momentum modes, general scattering-theory considerations lead 
to the Lehmann spectral decomposition \cite{bj} which implies that the 
wavefunction renormalization constant $Z_h$ (in $h_B(x) = \sqrt{Z_h} h_R(x)$) 
must satisfy $0 < Z_h \le 1$, with $Z_h \to 1$ in the continuum limit if the 
theory is `trivial.'   However, these well-known arguments place no 
constraint on $Z_{\phi}$, since there is no scattering theory for a 
zero-momentum mode --- the incident particles would never reach each other.  
Instead, $Z_{\phi}$ is fixed by the requirement that 
\BE
\label{concon}
\left. \frac{ d^2 V_{\rm eff}}{d \phi_R^2} \right|_{\phi_R=v_R} = M_h^2.
\EE
Although this is a familiar renormalization condition, the context here 
may be less familiar, so we would like to carefully explain its physical 
meaning in the `particle-gas' language.  

     First, consider a slight perturbation of the symmetric vacuum state 
(``empty box'').  We add a very small density $n$ of phions, each with 
zero 3-momentum.  The energy density is now:
\BE
{\cal E}(n) = 0 + n m + {\cal O}(n^2 \ln n),
\EE
where the first term is the energy of the unperturbed vacuum state (zero); 
the second term is the rest-mass cost of introducing $\num$ particles, divided 
by the volume; and the third term is negligible if we consider a sufficiently 
tiny density $n$.  Thus, we obviously have the relation 
\BE
\label{enm}
\left. \frac{ \partial {\cal E}}{\partial n} \right|_{n=0} = m.
\EE
The equivalent in field language follows from our previous relation 
\BE
\label{nphi}
n = \half m \phi_B^2
\EE
and is given by 
\BE 
{\cal E}(\phi_B) \equiv V_{\rm eff}(\phi_B) = 0 + \half m^2 \phi_B^2 + 
{\cal O}(\phi_B^4 \ln \phi_B^2),
\EE
so that 
\BE
\label{d2vb}
\left. \frac{ d^2 V_{\rm eff}}{d \phi_B^2} \right|_{\phi_B=0} 
= m^2.
\EE

     Now, let us consider a slight perturbation of the broken-symmetry 
vacuum (the box filled with a spontaneously-generated condensate).  Before 
we perturb it, this state has a density $n_v$ of phions, where $n_v$ is a 
(local) minimum of ${\cal E}(n)$.  From (\ref{nphi}) we have the translation 
$n_v = \half m v_B^2$.  This vacuum state, though complicated in terms 
of phions, is simple in terms of the `phonon' excitations corresponding
to Higgs bosons:  by definition it is just the state with no phonons.  
We now perturb it by adding a small density $n'$ of phonons, each with 
negligibly small 3-momentum.  (As noted above, phonons with {\it zero} 
momentum, created by ``$b_0^\dagger$,'' are undefined; here we are effectively 
defining them by continuity.)  The energy density of the perturbed state is 
then 
\BE
\label{enp}
{\cal E}(n') = {\cal E}(n'=0) + n' M_h + \ldots,
\EE
where the first term is the energy density of the unpertubed state 
($={\cal E}(n_v)$); the second term is the rest-mass cost of the added 
phonons; and any other terms from phonon interactions are negligible if 
$n'$ is small enough.  Thus, paralleling (\ref{enm}) we have 
\BE
\label{phmass}
\left. \frac{ \partial {\cal E}}{\partial n'} \right|_{n'=0} = M_h.
\EE
It is now natural to define a phonon field whose constant part, $f$, is 
related to the phonon density $n'$ by the analog of (\ref{nphi}), namely 
\BE
\label{npf}
n' \equiv \half M_h f^2.  
\EE
The ``renormalized field'' $\phi_R$ is simply this $f$ plus a constant.  
A constant must be added if we want to have $\phi_R$ proportional to $\phi_B$.  
Since, by definition, $f=0$ when $\phi_B=v_B$, we need
\BE
\phi_R \equiv f +v_R
\EE
with 
\BE 
\frac{v_R}{v_B} = \frac{\phi_R}{\phi_B} \equiv \frac{1}{Z_{\phi}^{1/2}}.
\EE
Now we can eliminate $f$ in favour of $\phi_R$ and re-write (\ref{npf}) 
as
\BE
\label{np}
n' = \half M_h (\phi_R - v_R)^2.
\EE
Hence, (\ref{enp}) can be re-written in field language as 
\BE
{\cal E}(\phi_R) \equiv V_{\rm eff}(\phi_R) =  V_{\rm eff}(\phi_R=v_R) 
+ \half M_h^2 (\phi_R - v_R)^2 + \ldots .
\EE
The crucial condition, Eq. (\ref{concon}), follows directly from this.  
It just says that the phonon mass is, self-consistently, $M_h$.  It is also, 
of course, the broken-vacuum counterpart of Eq. (\ref{d2vb}) for the symmetric 
vacuum.  

     The moral of this story is that the constant field $\phi_R - v_R$ is 
related to phonon density $n'$ in the same fashion that $\phi_B$ is related 
to the phion density $n$.  Note that there is a duality under
\BE
\mbox{\rm phions} \leftrightarrow \mbox {\rm phonons}, \quad \quad 
\phi_B \leftrightarrow \phi_R - v_R, \quad \quad 
n \leftrightarrow  n', \quad \quad 
Z_{\phi} \leftrightarrow  Z_{\phi}^{-1}.
\EE
Physically, this means that, we may choose either phion or phonon 
degrees of freedom to describe the theory.  Small excitations about the 
spontaneously broken vacuum are easily described in terms of phonons, but 
are complicated in terms of phions.  Likewise, small excitations about the 
symmetric vacuum are easily described in terms of phions, but are complicated 
in terms of phonons.  

\setcounter{equation}{0}
\section{Renormalized form of $V_{\rm eff}$ }

     Using the renormalized field $\phi_R = Z_{\phi}^{-1/2} \phi_B$ 
introduced in the last section we can write the effective potential in 
manifestly finite form.  It is convenient to define a finite parameter 
$\zeta$ in terms of the physical mass and the renormalized vacuum 
expectation value:  
\BE
\label{zeta}
\zeta \equiv \frac{M_h^2}{8 \pi^2 v_R^2}.
\EE
Using $M_h^2= \half \lambda v_B^2$ and $Z_{\phi}= v_B^2/v_R^2$, 
one then has 
\BE
\label{phirb}
               Z_{\phi}\equiv \frac{16 \pi^2}{\lambda} \zeta ,
\EE
so that $Z_{\phi}$ is of order $1/\lambda$, and hence of order $\ln \Lambda$.  
Imposing the condition 
\BE
\label{concon1}
\left. \frac{ d^2 V_{\rm eff}}{d \phi_R^2} \right|_{\phi_R=v_R} = M_h^2 ,
\EE
on the potential in Eq. (\ref{equiv}) one finds
\BE
\label{zeta2}
      m^2=\frac{\lambda}{16 \pi^2} 
           \left( {{\zeta-1}\over{2\zeta}} \right) M^2_h ,
\EE
leading to the final form of $V_{\rm eff}$:  
\BE
\label{vrenor}
%
V_{\rm eff}(\phi_R)=
\pi^2 \zeta(\zeta-1)\phi^2_R~(2 v_R^2-\phi^2_R) + 
\pi^2\zeta^2\phi^4_R~\left(\ln{{\phi^2_R}\over{v^2_R}}-{{1}\over{2}}\right) .
\EE

     The two independent quantities $\zeta$ and $v^2_R$ provide an 
{\it intrinsic} parametrization of the effective potential and replace 
the two bare parameters ($m^2$, $\lambda$) of the original Hamiltonian.  
Both the mass and the vacuum energy density can be expressed in terms of 
these parameters:
\BE 
M_h^2 = 8 \pi^2 \zeta v_R^2,
\EE
\BE
\label{minimo}
V_{\rm eff}(\phi_R=\pm v_R)=-\frac{\pi^2}{2} \zeta(2-\zeta) v_R^4.  
\EE
The values $\phi_R=\pm v_R$ are local minima of the effective potential 
for all positive values of $\zeta$.  However, only for $\zeta <2$ do 
these non-trivial minima have lower energy density than the symmetric vacuum.  
Thus, the symmetry-breaking phase transition occurs at $\zeta=2$ 
(corresponding to the value $m^2=m^2_c$).  At $\zeta=1$ one reaches the 
massless case (or `Coleman-Weinberg regime' \cite{cw}) where $m^2=0$.  
In the range $2\geq \zeta \geq 1$, where spontaneous symmetry breaking 
happens even for a positive physical phion mass $m$, the Higgs mass 
lies in the range 
\BE
\label{upper}
        4 \pi v_R \geq M_h \geq 2\sqrt{2} \pi v_R .
\EE
Finally, the range $ 1 > \zeta > 0$ corresponds to negative values of $m^2$ 
(`tachionic phions') where $M^2_h$ can become arbitrarily small in units of 
$v^2_R$.  In the extreme case $\zeta \to 0$ one recovers the 
classical-potential results.  

    It is important to stress that the final, renormalized result for the 
effective potential, (\ref{vrenor}), has a more general validity than the 
specific bare expression Eq. (\ref{veffeq}).  For instance, the Gaussian 
approximation generates a different bare expression, but leads to exactly 
the same renormalized result \cite{bran}.  The point is that, once the 
bare result is re-expressed in terms of the renormalized field through the 
condition (\ref{concon1}), the coupling $\lambda$ no longer appears.  Thus, 
the final result is the same in {\it any} approximation related to 
(\ref{veffeq}) by a replacement of the nominal coupling constant $\lambda$ 
by some effective coupling $\tilde{\lambda}$ \cite{bran,csz,agodi,rit2}.
Furthermore, any sort of effective coupling, $\tilde{\lambda}$, defined 
by summing some class of 4-point diagrams, is naturally of the same size 
as the original $\lambda$, if the latter has a size ${\cal O}(1/\ln \Lambda)$.  
The reason is that in Feynman graphs each loop generates at most a 
$\ln \Lambda$ factor \cite{fntequad}.  Thus, when each vertex has a factor 
$\lambda \sim {\cal O}(1/\ln \Lambda)$, any 4-point diagram, of arbitrary 
complexity, is at most of order $1/\ln \Lambda$.  That is, 
$\tilde{\lambda}= c \lambda$, where the finite number $c$ depends on 
precisely which class of diagrams have been taken into account.  In the same 
manner one can show that any contribution to $V_{\rm eff}$ not absorbed 
by the $\lambda \to \tilde{\lambda}$ replacement will be suppressed by one 
or more powers of $1/\ln \Lambda$ \cite{primer}.  This counting argument 
accords with our argument that, if the theory is `trivial,' it must be 
possible to reabsorb all interaction effects into suitable redefinitions 
of the classical energy density term, 
$(\lambda/4!)\phi^4_B \to (\tilde{\lambda}/4!)\phi^4_B$, and of the mass 
$M^2(\phi_B)=\half \lambda\phi^2_B \to \half \tilde{\lambda}\phi^2_B$,  
which governs the zero-point energy contribution from the free-field 
fluctuations of the shifted field (see Fig. 3).

\setcounter{equation}{0}
\section{Summary: a hierarchy of length scales }

   We have reconsidered the symmetry-breaking phase transition in 
$\lambda\Phi^4$ theory from a fresh perspective.  The physics can be 
understood intuitively in terms of actual particles and their interactions.  
We have shown that the energy density ${\cal E}(n)$, for a given particle 
density $n$, consists of $n$, $n^2$, and $n^2 \ln n$ terms arising from 
rest masses, short-range 2-body repulsions, and long-range 2-body attractions, 
respectively.  The crucial $\ln n$ factor arises because the $-1/r^3$ 
attraction is so long range that it generates a logarithmic infrared 
divergence that is tamed only by screening from the background density.  
The translation $n= \half m \phi^2$ converts ${\cal E}(n)$ to the 
field-theoretic effective potential $V_{\rm eff}(\phi)$.  

   The resulting form of $V_{\rm eff}(\phi)$ --- a sum of $\phi^2$, 
$\phi^4$, and $\phi^4 \ln \phi^2$ terms --- is exactly what one should 
expect in a `trivial' theory, as we argued in the introduction.  
Moreover, we self-consistently find that the theory {\it is} `trivial' 
because the scattering length $a$ tends to zero in the continuum limit.  

    Despite `triviality' --- in fact, {\it because of} `triviality' --- 
a rich hierarchy of length scales emerges.  This hierarchy is summarized 
in the figure below in terms of the small parameter 
$\epsilon \equiv  1/\ln (\Lambda/M_h)$, which tends to zero in the continuum 
limit, $\Lambda \to \infty$.  

\begin{tabular}{ccccc}
\hspace*{12mm} & \hspace*{2.1cm} & \hspace*{2.4cm} & 
\hspace*{3.0cm} & \hspace*{3.5cm} \\
      & $v_B^{-1}$  &     &  $v_R^{-1}$  &    
\vspace*{1mm} \\
\hline 
\end{tabular}

\vspace*{-9mm} 

\begin{tabular}{ccccc}
\hspace*{4mm} & \hspace*{2.1cm} & \hspace*{2.4cm} & 
\hspace*{3.0cm} & \hspace*{3.5cm} \\
$|$   & $|$   &   $|$          & $|$         &  $|$    \\
$r_0$   & $a$   & $d$            &  $\xi_h$    &   $\xi$  \\
$(\Lambda^{-1})$  & $\left(\frac{\lambda}{8\pi m}\right)$  
& $(n_v^{-1/3})$  &  $(M_h^{-1})$  &   $(m^{-1})$ 
\vspace*{5mm} \\
${\rm e}^{-1/\epsilon}$ & $\epsilon^{1/2}$  & $\epsilon^{1/6}$ & $1$ & 
$\epsilon^{-1/2}$  
\end{tabular}

\vspace*{5mm}
\hspace*{1.5cm} {\bf Figure 4}:  
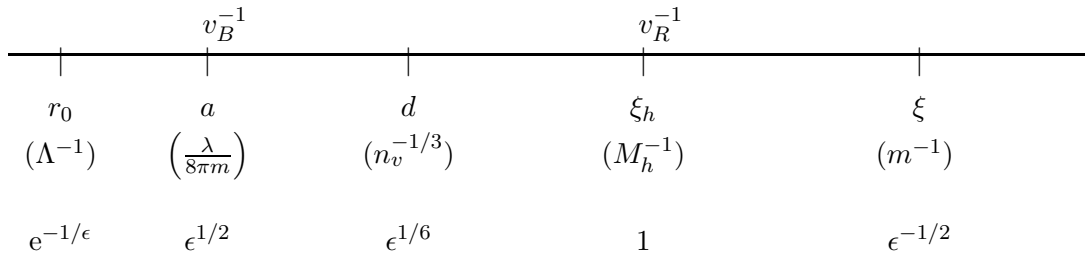
Schematic representation of the length-scale hierarchy. \\
\hspace*{4cm} $\epsilon \equiv 1/\ln(\Lambda/M_h)$.  

\vspace*{5mm}

    The `unit of length' here is the correlation length of the broken phase, 
$\xi_h \equiv 1/M_h$, the inverse of the Higgs mass.  The phion Compton 
wavelength, $\xi \equiv 1/m$, is much longer since, at or near the phase 
transition, the phion mass $m$ becomes infinitesimal;  
$m^2/M^2_h \sim \epsilon$.  At the same time, the phion-phion coupling 
becomes infinitesimal: $\lambda \sim \epsilon$.  The phion-phion cross 
section due to short-range repulsion is proportional to the square of the 
scattering length $a(E) = \frac{\lambda}{8 \pi E}$.  Even for the lowest 
phion energy, $E=m$ the scattering length $a = \frac{\lambda}{8 \pi m}$ 
vanishes like $\epsilon^{1/2}$ in length units set by $\xi_h \equiv 1/M_h$.  

     The phion density at the minimum of the effective potential $\phi_B=v_B$ 
is $n_v={{1}\over{2}}m v^2_B$, which is very large, 
${\cal O}(\epsilon^{-1/2})$, in physical units.  
Hence, the average spacing between two phions in the condensate, 
$d \equiv n_v^{-1/3}$, is tiny compared with $\xi_h$:
${{d}\over{\xi_h}} \sim \epsilon^{1/6}$.  It is because there is such a 
high density of phions that their tiny interactions produce a finite effect 
on the energy density.  Nevertheless, the phions in the condensate are very 
dilute because $n_v a^3 \sim \epsilon$.  In other words, the average spacing 
between phions is much, much larger than their interaction size: 
$d/a \sim \epsilon^{-1/3}$.  

     Upon translating back to field language, a crucial element of this 
picture is the large re-scaling of the zero-momentum (spacetime constant) 
part of the field.  The normalizations of $\phi_B$ and $\phi_R$ are set by 
the conditions 
\BE
\label{d2vb2}
\left. \frac{ d^2 V_{\rm eff}}{d \phi_B^2} \right|_{\phi_B=0} \equiv  m^2, 
\quad \quad \quad 
\left. \frac{ d^2 V_{\rm eff}}{d \phi_R^2} \right|_{\phi_R=v_R} \equiv M_h^2.
\EE
Since the theory is ``nearly'' a massless, free theory, $V_{\rm eff}$ 
is a very flat function of $\phi_B$, and so the re-scaling factor $Z_{\phi}$
in $\phi_B^2 = Z_{\phi} \phi_R^2$ is large, ${\cal O}(1/\epsilon)$.  
Thus, the length scale $v_B^{-1}$ is of order $\epsilon^{1/2}$, comparable 
to the scattering length $a$, while the length scale $v_R^{-1}$ 
is finite, comparable to $M_h^{-1}$.  

    The existence of a $Z_{\phi}$, distinct from the wavefunction 
renormalization, $Z_h$, of the finite-momentum modes, is now supported by 
some direct lattice evidence \cite{ising}.  A Monte-Carlo simulation of the 
Ising limit of $\LP$ theory was used to measure (i) the zero-momentum 
susceptibility $\chi$: 
\BE
\chi^{-1}=
\left. \frac{ d^2 V_{\rm eff}}{d \phi_B^2} \right|_{\phi_B=v_B} 
\equiv {{M^2_h}\over{ Z_{\phi}}}, 
\EE
and (ii) the propagator of the shifted field (at Euclidean momenta $p \neq 0$). 
The latter data was fitted to the form 
\BE
              G(p)= {{Z_h}\over{ p^2 + M^2_h}}
\EE 
to obtain the mass and wavefunction-renormalization constant $Z_h$.  The 
resulting $Z_h$ is slightly less than one, and seems to approach unity as 
the continuum limit is approached, consistent with the expected `triviality' 
of the field $h(x)$ in the continuum limit.  However, the $Z_{\phi}$ 
extracted from the susceptibility is clearly different.  It shows a rapid 
increase above unity, and the trend is consistent with it diverging in the 
continuum limit.  

    This evidence, and the earlier lattice results for $V_{\rm eff}$ 
\cite{agodi,cea}, provide objective support for our picture of symmetry 
breaking in $\LP$ theory.  As we have tried to show in this paper, the 
picture also has an appealing and very physical interpretation.  

     We would like to close with some speculations about some possible 
implications of these ideas in relation to gravity.  Unlike other fields, 
which couple to gravity only via the $\sqrt{{\rm det} g}$ factor, a scalar 
field has a direct coupling to gravity through an $R \Phi^2$ term in the 
Lagrangian, where $R$ is the curvature scalar.  For this reason, it has been 
proposed that Einstein gravity could emerge from spontaneous symmetry breaking 
\cite{fuji,zee,adler}.  The Newton constant, just like the Fermi constant, 
would then arise from the vacuum expectation value of a scalar field.  
It was suggested by van der Bij \cite{bij} that the scalar field inducing 
gravity could be {\it the same} scalar field responsible for electroweak 
symmetry breaking.  The problem, though, is to understand the origin of the 
large re-scaling factor $\eta \sim 10^{34}$ needed in the coupling 
$\eta R \langle\Phi\rangle^2$ to obtain the Planck scale $\sim 10^{19}$ GeV 
from the Fermi scale $\sim 10^2$ GeV.  Our results offer a possible solution 
to this puzzle.  If we identify the Fermi scale with the physical vacuum 
field $v_R$, it is naturally infinitesimal with respect to the Planck scale, 
if we identify the latter with the bare condensate $v_B$.  (This means that 
gravity must probe the scalar condensate at a much deeper level than 
electroweak interactions.)   In this scenario we would need to back off from 
literally taking the $\Lambda \to \infty$ limit; instead we would need 
$\eta\equiv Z_\phi\sim \ln(\Lambda/M_h)$ to be large but finite, 
$\sim 10^{34}$.  Note that this implies a cutoff $\Lambda$ that is enormously 
larger than the Planck mass \cite{fntecutoff}.

     One might wonder if gravity can still be neglected in discussing the 
phion dynamics.  The formation of the phion condensate hinges on the very 
weak, long-range $-1/r^3$ potential, while gravity produces another weak 
--- and even longer-range --- interaction between phions.  However, all is 
well provided that $Gm^2/r$ is much less than ${\cal A}/r^3$ even for 
$r \sim r_{\rm max}$.  A little algebra shows that this condition is indeed 
satisfied, because the ratio $M_{\rm Planck}^2/M_h^2$ (of order 
$\ln \Lambda$ in the above scenario) is much, much greater than unity.  

     A related issue is the `inflaton' scalar field invoked in inflationary 
models of cosmology.  The extraordinary fine-tuning of the scalar 
self-coupling needed to obtain a very slow roll-over from the symmetric to 
the broken vacuum \cite{kolb} has led to the conclusion that {\it ``... the 
inflaton cannot be the Higgs field as had been originally hoped''} 
\cite{guth}.  However, the $\phi_B$, $\phi_R$ distinction in our picture 
offers a natural way out of the difficulty.  If gravity couples to the 
{\it bare} condensate, as postulated above, then it indeed sees an 
extremely flat effective potential.  In our picture, a finite vacuum energy 
and a finite Higgs mass coexist with an infinitesimal slope of the effective 
potential (parametrized in terms of the bare vacuum field).  

    In our approach the natural order of magnitude relation is $M_h=O(v_R)$,
with $m^2/M_h^2 \sim 1/Z_\phi$.  In the scenario above, where 
$Z_\phi \sim 10^{34}$, we might expect a phion mass of order 
$m \sim 10^{-4}-10^{-5}$ eV or smaller.  Possibly, due to mixing effects 
with the graviton, this could produce deviations from the pure $1/r$ 
gravitational potential at the millimeter scale \cite{dimo} that can be 
tested in the next generation of precise `fifth-force' experiments 
\cite{price}.

\begin{center}
{\bf Acknowledgements}
\end{center}

    This work was supported in part by the Department of Energy under Grant No. 
DE-FG05-92ER41031.


\begin{thebibliography} {99}
\bibitem{triviality}
  We shall consistently use the word `trivial' (in quotation marks) 
in the technical sense, meaning that all scattering amplitudes vanish 
when the continuum limit is taken \cite{book}.  One should not confuse 
this with the everyday meaning of `easy' or `uninteresting.'
\bibitem{massren}
More generally we mean this notation as a shorthand indication that any 
necessary mass-renormalization has been done; that is, a $\Phi^2$ and a 
$\Phi$-independent term are subtracted from $H$ such that the symmetric 
vacuum has zero energy, $\langle 0|H|0\rangle = 0$, and such that $m$ 
corresponds to the physical mass of the phion particle.
\bibitem{book2}
The result just deduced, that the ratio $M_h/v_B$ vanishes in the limit 
$\Lambda \to \infty$, agrees with rigorous quantum field theoretical 
arguments.  See chapter 15 of Ref. \cite{book}.
\bibitem{fntequad} 
We ignore quadratic-divergent contributions since these are absorbed by 
mass renormalization.  Alternatively, we may re-cast the argument using 
dimensional regularization.
\bibitem{fntecutoff}
However, $\Lambda$ may not represent an actual mass scale; it might be that 
spacetime foam, or extra compact dimensions, etc., produce something like 
dimensional regularization, with an ``effective spacetime dimension'' $d$ 
close to 4.  Then $4-d$ would play the role of 
$\epsilon \equiv 1/\ln(\Lambda/M_h)$.
\end{thebibliography}
\end{document}